\documentclass[12pt,preprint]{aastex}

\shorttitle{Magnetic Reconnection of A Flare} \shortauthors{Li \&
Zhang}

\begin{document}

\title{Observations of The Magnetic Reconnection Signature of An M2 Flare on 2000 March 23}

\author{Leping Li and Jun Zhang }

\affil{Key Laborary of Solar Activity, National Astronomical
Observatories, Chinese Academy of Sciences, Beijing 100012, China;
lepingli;zjun@ourstar.bao.ac.cn}

\begin{abstract}

Multi-wavelength observations of an M 2.0 flare event on 2000 March
23 in NOAA active region 8910 provide us a good chance to study the
detailed structure and dynamics of the magnetic reconnection region.
In the process of the flare, extreme ultraviolet (EUV) loops
displayed two times of sideward motions upon a loop-top hard X-ray
source with average velocities of 75 and 25.6 km $s^{-1}$,
respectively. Meanwhile part of the loops disappeared and new
post-flare loops formed. We consider these two motions to be the
observational evidence of reconnection inflow, and find an X-shaped
structure upon the post-flare loops during the period of the second
motion. Two separations of the flare ribbons are associated with
these two sideward motions, with average velocities of 3.3 and 1.3
km $s^{-1}$, separately. The sideward motions of the EUV loops and
the separations of the flare ribbons are temporally consistent with
two peaks of the X-ray flux. This indicates that there are two times
of magnetic reconnection in the process of the flare. Using the
observation of photospheric magnetic field, the velocities of the
sideward motions and the separations, we deduce the corresponding
coronal magnetic field strength to be about 13.2-15.2 G, and
estimate the reconnection rates to be 0.05 and 0.02 for these two
magnetic reconnection process, respectively. Besides the sideward
motions of EUV loops and the separations of flare ribbons, we also
observe motions of bright points upward and downward along the EUV
loops with velocities ranging from 45.4 to 556.7 km $s^{-1}$, which
are thought to be the plasmoids accelerated in the current sheet and
ejected upward and downward when magnetic reconnection occurs and
energy releases. A cloud of bright material flowing outward from the
loop-top hard X-ray source with an average velocity of 51 km
$s^{-1}$ in the process of the flare may be accelerated by the
tension force of the newly reconnected magnetic field lines. All the
observations can be explained by schematic diagrams of magnetic
reconnection.

\end{abstract}


\section{Introduction}

Solar flares are one of the most spectacular phenomena in solar
physics. They are sudden brightening in the solar atmosphere, and
consist of a number of components including post-flare loops
\citep{for83,li09}, ribbons \citep{iso02,din03,iso05,tem07},
arches \citep{mar88,tri06}, remote patches \citep{tan82,wan05},
surges \citep{roy73,jia07}, erupting filaments
\citep{gop89,zha01b,jia06}, and other expanding coronal features
\citep{mar89,wan92,zha01a,zha07}. They have been studied
morphologically from direct images
\citep[e.g.][]{kru00,fle01,ji06} and spectroscopically from
spectrograms \citep[e.g.][]{moo76,cow73,gri05} at different
wavelength regions. The theories for solar flares have also been
reviewed by many people, such as \citet{par63}, \citet{swe69},
\citet{van76}, \citet{pri76}, \citet{for03}, and \citet{gri06}.

Solar flares are now thought to be caused by magnetic
reconnection-the reorganization caused by local diffusion of
anti-parallel magnetic field lines in a certain local point in the
corona. The tension force of the reconnected magnetic field lines
then accelerates the plasma out of the dissipation point. Because of
this outflow, the ambient plasma is drawn in. The inflowing plasma
carries the surrounding magnetic field lines into the dissipating
point. These magnetic field lines continue the reconnection cycle.
Therefore, the magnetic energy stored near the dissipation point is
released to become the thermal and bulk-flow energy of plasma
\citep{yok01}. The evidence of magnetic reconnection found by space
observations includes the cusp-shaped post-flare loops
\citep{tsu92}, the loop-top hard X-ray source \citep{mas94}, the
reconnection inflow \citep{yok01,lin05,nar06}, downflows above
post-flare loops \citep{mck99,inn03,asa04}, plasmoid ejections
\citep{shi95,ohy97,ohy98}, etc. The magnetic reconnection model
proposed by \citet{car64}, \citet{stu66}, \citet{hir74}, and
\citet{kop76} (the CSHKP model) suggests that magnetic field lines
successively reconnect in the corona. This model explains several
well-known features of solar flares, such as the growth of flare
loops with a cusp-shaped structure and the formation of the
H$\alpha$ two-ribbon structures at their footpoints. In recent
decades, this model has been further extended
\citep[e.g.][]{moo01,yoko01,pri02,lin04}

In this paper, we analysis a flare event to investigate the detailed
structure and dynamics of the reconnection region. We show the
observational data in section 2, and the corresponding results in
section 3. Conclusions and brief discussion are presented in section
4.

\section{Observations}

On 2000 March 23, an M 2.0 flare occurred near the solar limb (N15
W69) in NOAA active region (AR) 8910. This flare started at 11:32
UT and ended at 12:30 UT, with a peak at 12:14 UT. It was observed
by several satellites including: \emph{Transition Region and
Coronal Explorer} \citep[\emph{TRACE};][]{han99}, \emph{Solar and
Heliospheric Observatory} (\emph{SOHO}), \emph{YOHKOH} and
\emph{GOES}. In this paper, we use \emph{TRACE} 195 \AA~images,
with 10 s temporal resolution and 1\arcsec~spatial resolution, to
study the dynamics of the extreme ultraviolet (EUV) loops during
the flare, and 1600 \AA~data, with 1\arcsec~spatial resolution and
1 minute temporal resolution, to research the kinetics of the
flare ribbons. The evolution of the magnetic fields and the
sunspots in the source region of the flare is studied using
\emph{SOHO}/\emph{Michelson Doppler Imager}
\citep[\emph{MDI};][]{sch95} magnetograms and \emph{TRACE} WL
images. We also employ observations of \emph{YOHKOH/Hard X-ray
Telescope} \citep[\emph{HXT};][]{kos91} and \emph{GOES} to explore
the X-ray variation of the event.

\section{Results}

The general information of this event is exhibited in Fig. 1.
Figure 1a shows the longitudinal magnetogram of the source AR of
the flare observed by \emph{SOHO/MDI}, and indicates that this AR
has a mixture of polarities and further complicated magnetic
neutral lines. In order to make sure the polarities of the
magnetic fields without the limb effect, we investigate all the
observations of \emph{SOHO/MDI} from 2000 March 20, when the AR
was in the center of the solar disk, to 2000 March 23, when the
flare occurred, and find that all the polarities of the magnetic
fields associated with the flare are true, denoted by P1, P2, P3
and P4 for positive ones, while N1, for negative. The continuum
intensity observed by \emph{TRACE} WL is displayed in Fig. 1b.
Figure 1c exhibits \emph{TRACE} 1600 \AA~image. This flare is a
complicated one with several flare ribbons shown as FR1, FR2 and
FR3. We overlay the magnetic fields shown in Fig. 1a as white and
black contours on Fig. 1c, and find that the southern flare ribbon
FR1, a simple one, foots around N1, while the northern flare
ribbons FR2 and FR3, more complicated, around P2, P3 and P4.
Figures 1d-1f show the time sequence of \emph{TRACE} 195
\AA~images. Before the occurrence of the flare, there were a set
of EUV loops in the AR marked as L1 in Fig. 1d. Comparing Fig. 1d
with Figs. 1a-1c, we notice that L1 connects P1 and N1.

\subsection{Loop dynamics}

From about 11:35 UT, a sideward motion of L1 appeared from southwest
to northeast, i.e. along the solid line BA shown in Fig. 1d,
meanwhile L1 partly disappeared. After several minutes of this
motion, new post-flare loops denoted by PFLs1 in Fig. 1e appeared.
Comparing Fig. 1e with Fig. 1c, we find that the southern leg of
PFLs1 foots at FR1, while the northern one, at FR2. The contours on
Fig. 1d represent the hard X-ray emissions in the L energy band
(14-23 keV) observed by \emph{Yohkoh/HXT} at 11:35 UT (see the left
vertical arrow in Fig. 4d). The majority of the hard X-ray sources,
which can be considered as footpoint sources, are co-spatial with
the flare ribbons, and a source marked as LTS1, which also appears
in the M1 energy band (23-33 keV) image, a loop-top one. The height
of LTS1 to footpoint sources (see the two-head arrows in Fig. 1d) is
about 20 Mm. As this flare is a limb event, the projection effect
must be considered. We assume that the line along the two-head
arrows is vertical to the local photosphere, and its heliographic
position is the same to the AR (N15 W69). Then the height
$H_{corrected}$ after correction
\begin{equation}
H_{corrected}=\frac{H_{measured}}{\sqrt{(\sin15)^2+(\cos15)^2(\sin69)^2}}
\end{equation} is 22.7 Mm, where $H_{measured}$ is the measured
value. The physical parameters mentioned below, e.g. the length of
the current sheet, which are similar to the height of LTS1, are
corrected using the same method. The loops marked as L11 in Fig.
1e, which was part of L1, moved toward the northeast from 11:43
UT, and disappeared at 11:44 UT.

From about 11:49 UT, another sideward motion of L1 was detected at a
higher position from southeast to northwest, i.e. along the solid
line DC (shown in Fig. 1f), and another set of post-flare loops
marked as PFLs2 in Fig. 1f appeared. Comparing Fig. 1f with Fig. 1c,
we note that the southern leg of PFLs2 also foots at FR1, but the
northern one, at FR3. By overlaying the contour of a hard X-ray
image observed by \emph{Yohkoh/HXT} at 12:06 UT (see the right
vertical arrow in Fig. 4d) on Fig. 1f, we find that there is also a
loop-top source marked as LTS2, with a height of 30 Mm (34 Mm after
correction). From Figs. 1d-1f, we uncover that L1 firstly undergoes
sideward motion and partly disappears, then post-flare loops form.
Furthermore, the hard X-ray sources (LTS1 and LTS2) locate upon the
top of post-flare loops, and under the region where the loops show
maximum sideward motion velocities.

The second sideward motion of L1 lasts for a longer time (22
minutes) than the first one (6 minutes), and the flare is more
powerful in this period, so we study it in detail.  Figure 2
displays the time sequence of TRACE 195 \AA~images showing the
second sideward motion. The dotted lines in Figs. 2a-2b represent
the EUV loops at 11:54:24 UT. From these two figures, we can see
clearly the sideward motion (denoted by a white arrow in Fig. 2b).
When two approaching loops met, the motion of these loops stopped,
but there were still some neighboring loops moved toward the meeting
region continuously. The distance between these two EUV loops is
almost constant with a value of 0.3-0.7 Mm (shown by two black
arrows in Fig. 2c). We assume that the direction of the sideward
motions is perpendicular to the line of sight, so the projection
effect to the distance can be neglected. The physical parameters
mentioned below, e.g. the velocities of sideward motions, which are
similar to the distance, are not necessary to be corrected for the
projection effect. The dashed lines in Fig. 2c, the outer edges of
the EUV loops, show an X-shaped structure. From 11:58:11 UT, the
second set of post-flare loops (denoted as PFLs2 in Fig. 2d)
appeared with a cusp-shaped structure (see CUSP in Fig. 2d) riding
on it. Before 11:59 UT, the shape of L1 was smooth, then an abrupt
break occurred accompanying with the appearance of a brightening
point ``B" (see Fig. 2d). After the break, several other bright
points, such as ``BP" marked in Fig. 2e, appeared above the ``B"
point, and propagated upward along L1 (see the white solid arrow in
Fig. 2e). The EUV loops L (arrowed in Fig. 2d) under the ``B" point
disappeared after several minutes of the break, only the post-flare
loops PFLs2 and the cusp-shaped structure CUSP left (see Fig. 2f).
We estimate the length of the loops between ``B" point and CUSP to
be 12-15 Mm (shown by the two-head solid arrows in Fig. 2d), that is
13.6-17 Mm after correction.

In order to quantificationally study the sideward motions of the
EUV loops, we make a time slice along the moving directions of L1,
and show them in Figs. 3a and 3b. Figure 3a presents the time
evolution of the one-dimension distribution of EUV intensity of
the loops along BA from 11:34 to 11:45 UT. In this figure, the
left arrows marked as F1 show the first sideward motion with an
average velocity of 75 km $s^{-1}$. The dotted line indicates that
part of L1 first moved along BA, then returned after several
minutes. Furthermore, a small portion of the returning loops also
displayed a sideward motion along BA, as denoted by L11 (see also
Fig. 1e), with an average velocity of 70 km $s^{-1}$. Figure 3b is
the time slice of the second main sideward motion of L1 along DC
from 11:45 to 12:12 UT. A clear merging pattern (see also in Fig.
2) arrowed as F2 can be seen with an average velocity of 25.6 km
$s^{-1}$.

\subsection{Flare ribbon kinetics and X-ray flux properties}

In the CSHKP model, the reconnection points move upward,
therefore, newly reconnected field lines have their footpoints
further out than that of the field lines which have already
reconnected, which leads us to recognize the ``apparent"
separation motion of the flare ribbons \citep{asa04}. We exhibit
the time evolution of the one-dimensional distribution of 1600
\AA~intensity of FR1 along the solid line FE (see Fig. 1c) from
11:34 to 12:12 UT in Fig. 3c. There are two clear separations
shown as S1 and S2 with the average velocities of 3.3 and 1.3 km
$s^{-1}$, respectively. Between these two separations, FR1 became
too weak to be observed.

Figure 3d shows \emph{GOES}-10 1-8 \AA~soft X-ray flux (dashed
curve) of this flare, and indicates that there are two peaks (PS1
and PS2) of the flux at 11:37 and 12:14 UT, separately. The solid
curve shows the time integrated hard X-ray flux observed by
\emph{YOHKOH/HXT}. Unfortunately, there are no observations between
11:36 and 12:03 UT. As there is a good correlation between the time
derivative of soft X-ray flux and the hard X-ray one \citep{den93},
we use the time derivative (the dotted curves) of \emph{GOES} soft
X-ray flux to extrapolate the change of the hard X-ray one during
the observational gap. Comparing the hard X-ray flux (solid curves)
with the time derivative (dotted curves), we find that the peaks of
these two flux curves are similar. There are two peaks (PH1 and PH2)
which correspond with PS1 and PS2. The vertical dash-dotted lines
represent the beginning time of two sideward motions. From Fig. 3,
we notice that the two sideward motions (F1 and F2) of L1, the two
separations (S1 and S2) of FR1, and the two peaks (PS1 and PS2, PH1
and PH2) of X-ray flux are temporally consistent.

\subsection{Plasma ejections}

Besides the sideward motions of the EUV loops and the separations of
the flare ribbons, we also find plasma ejections in the process of
the flare. Immediately after the beginning of the sideward motion,
lots of bright points appeared and propagated upward and downward
along the loops. We show an example of a pair of bright points in
Figs. 4a-4c. These figures display the running difference images of
\emph{TRACE} 195 \AA~from 11:37:59 to 11:38:44 UT. From these
images, we find a black point (arrowed by UF) moving upward along
L1, with an average velocity of 221.2 km $s^{-1}$ (251.3 km $s^{-1}$
after correction), as well as another black point (denoted by DF),
downward, with an average velocity of 167.3 km $s^{-1}$, which is
190.1 km $s^{-1}$ after correction. The upward and downward motions
of bright points existed all the time during the flare. During the
process of the first sideward motion, the upward bright points
appeared always higher than the region where the loops displayed
maximum motion velocities, and the downward ones, under that region.
Both of them propagated along L1. In the process of the second
sideward motion, the upward bright points appeared upon the bright
point ``B" and moved along the right leg of the X-shaped structure,
while the downward ones, appeared under the cusp-shaped structure
and moved along the right leg of PFLs2. The area of the moving
bright points ranges from 1.5 to 23 $Mm^2$, with an average value of
9.1 $Mm^{2}$. The moving speeds of the bright points range from 40
to 490 km $s^{-1}$ (45.4-556.7 km $s^{-1}$ after correction), with
an average value of 172 km $s^{-1}$ (195.4 km $s^{-1}$ after
correction).

Moreover, we find an outflow of a bright cloud. Figures 4d-4f show a
series of \emph{TRACE} 195 \AA~running difference images. The cloud
of the bright material (arrowed by OF) went away upon the loop-top
hard X-ray source LTS2 with velocities of 25.4-79.4 km $s^{-1}$
(28.9-90.2 km $s^{-1}$ after correction) and an average value of
44.9 km $s^{-1}$ (51 km $s^{-1}$ after correction). In the
propagating process of the outflow, these bright material became
diffused, and disappeared after 20 minutes.

\section{CONCLUSIONS AND DISCUSSION}

In this paper, we analyze an M 2.0 flare on 2000 March 23 at N15
W69 for detail, and get the following results: 1. Long EUV loops
undergo two times of sideward motions and partly disappear,
subsequently two sets of post-flare loops form. 2. There are two
peaks in the X-ray flux of the flare. Each peak is temporally
consistent with a phase of the sideward motion of the EUV loops
and the separation of the flare ribbons. 3. Bright points eject
along the EUV loops upward and downward all the time during the
flare. 4. An outflow of a bright cloud moves away from the
loop-top hard X-ray source region.

It is well accepted that magnetic reconnection in solar corona
results in solar flares. \citet{mas94} suggested that magnetic
reconnection takes place around or above the loop-top hard X-ray
source. In our study, the sideward motions and disappearances of EUV
loops also take place above the loop-top hard X-ray sources (see
Figs. 1d and 1f), which may be consistent with that of Masuda et
al.. \citet{tsu92} observed cusp-shaped post-flare loops, and
suggested that an X-type or Y-type reconnection point wound be
formed at the top of the cusp. In the process of the second sideward
motion, we find an X-shaped structure above the cusp-shaped
post-flare loops (see Fig. 2) which is identical with the X-type
current sheet mentioned by Tsuneta et al., and the sideward motions
can be considered as reconnection inflows. As the two peaks of the
X-ray flux are relevant to the two times of sideward motions of
loops and separations of flare ribbons, we consider that there are
two magnetic reconnection process in this flare.

Although much evidence has been found to support the magnetic
reconnection mechanism, the slow and fast shocks predicted by
reconnection theories \citep{pet64,for83,uga87} have not yet been
identified. For the detailed observations of the flare, we have the
chance to search for the signature of shocks associated with
magnetic reconnection. \citet{shi03} once performed MHD simulations
of a giant arcade formation with a model of magnetic reconnection
coupled with heat conduction, and showed that the Y-shaped structure
was identified to correspond to the slow and fast shocks associated
with the magnetic reconnection. In this work, we study the physical
parameters at the similar positions and similar times of the
simulation of Shiota et al.. Figure 5a shows the distributions of
several physical quantities, e.g. temperature, emission measure and
brightness, along the white line GH (see Fig. 2c) at 11:58:11 UT.
The temperature and emission measure are calculated from a
wavelength pair (171 and 195 \AA) of two TRACE images. From Fig. 5a,
we can see a discontinuous region ``I" between two dash-dotted lines
which is similar to the simulated slow shock. It may be the
observational evidence of the slow shock associated with magnetic
reconnection. \citet{mas94} once suggested that the hard X-ray
source above the loop top in an impulsive flare may be a fast shock
created by the collision of a reconnection jet with the flare loop.
The hard X-ray source LTS2 above the top of post-flare loops and the
slightly brighter cross points of the X-shaped structure may be the
observational evidence of fast shocks. In order to confirm this, we
show the distributions of some physical quantities along the white
line IJ (see Fig. 2c) in Fig. 5b. There are two discontinuous
regions (``II" and ``III") between the dash-dotted lines. The
difference between these two regions may be caused by the different
local magnetic field configurations. By comparing our observations
with the simulations of Shiota et al., we find that these two
discontinuous regions may be identical with the fast shocks.
Therefore, the X-shaped structure may correspond to the slow and
fast MHD shocks associated with magnetic reconnection.

A similar process of reconnection inflow was reported by
\citet{yok01} for the event on 1999 March 18 with an inflow
velocity of 1.0-4.7 km $s^{-1}$. \citet{lin05} showed another
example of reconnection inflow by analyzing a flare event on 2003
November 18, and gave out the average velocities of 10.5-106 km
$s^{-1}$. \citet{nar06} statistically analyzed six reconnection
inflows in solar flares observed with \emph{SOHO/EIT}, and found
the inflow velocities were about 2.6-38 km $s^{-1}$. We use TRACE
data with higher spatial and temporal resolutions to get the
inflow velocities to be 75 and 25.6 km $s^{-1}$, more likely
consistent with Lin et al.. In the process of the flare, we also
observe two times of separations of flare ribbon. Using the
conservation of magnetic flux
\begin{equation} v_{inflow}B_{corona}=v_{foot}B_{photo},
\end{equation} we can estimate the coronal magnetic field strength
$B_{corona}$, where $B_{photo}$ and $v_{foot}$ are the
photospheric magnetic field strength and separation velocity of
the flare ribbons \citep{iso02}. From this equation, we get
$B_{corona}$ as
\begin{equation}
B_{corona}=B_{photo}\frac{v_{foot}}{v_{inflow}}.
\end{equation}
For the two times of magnetic reconnection, the ratios of the
separation velocities of flare ribbons to inflow velocities of EUV
loops are 0.044 and 0.051 respectively, one order of magnitude
smaller than that of \citet{nar06}. The average magnetic field
strength in the photosphere of this AR is about 300 G. Then
$B_{corona}$ is about 13.2 and 15.2 G during the process of the
two times of magnetic reconnection, respectively. The local
Alfv\'{e}n velocity $v_{A}$ is expected as
\begin{equation}
v_{A}=\frac{B_{corona}}{\sqrt{4\pi\rho}}=\frac{B_{corona}}{\sqrt{4\pi
m_{p}n_{p}}},
\end{equation}
where $m_{p}$=1.67$\times$10$^{-24}$ g is the proton mass and
$n_{p}$ is the proton number density outside the current sheet which
is 4$\times$10$^{8}$ $cm^{-3}$ \citep{yam02,iso02}. We use 14 G as
$B_{corona}$ to calculate $v_{A}$ and obtain it to be 1530 km
$s^{-1}$, then estimate the reconnection rate $M_{A}$
\begin{equation}
M_{A}=\frac{v_{inflow}}{v_{A}} \end{equation} to be 0.05 and 0.02 in
two process of magnetic reconnection, separately. They are bigger
than Yokoyama et al. (0.001-0.03) and Narukage \& Shibata
(0.001-0.07), and smaller than Lin et al. (0.01-0.23). During the
second sideward motion, we obtain the distance between the X-shaped
EUV loops to be 0.3-0.7 Mm. If the distance is the width of the
current sheet in this flare, it should be the upper limit, and is
two orders of magnitude smaller than that of \citet{lin07}. We
estimate the length of the current sheet to be the distance between
the top of the cusp-shaped structure and the ``B" point (see the
two-head solid arrows in Fig. 2c), and get a value of 13.6-17 Mm.
According to Sweet-Parker model, if the plasma compressibility is
neglected, the reconnection rate is given by the ratio of width to
length of the current sheet \citep{pri02}. So we calculate the
reconnection rate to be 0.02-0.05 which is similar with those values
(0.05 and 0.02) we estimated above. The height of the loop-top hard
X-ray source of two magnetic reconnections are 22.7 and 34 Mm which
may represent the height of the lower parts of the current sheet.
Both of these heights are much lower than that of \citet{yok01},
\citet{lin05} and \citet{nar06}.

In the process of magnetic reconnection, many bright points ejected
along the EUV loops upward and downward. These ejections may be
plasmoids accelerated in current sheet. \citet{mck99} examined the
super-arcade downflow motions sunward from the high corona with
speeds of 45-500 km $s^{-1}$. \citet{inn03} found that highly
blueshift features, which correspond to a Doppler velocity of up to
1000 km $s^{-1}$, were associated with the downflows using
\emph{SUMER} and \emph{TRACE} observations. \citet{asa04} found
downflows with velocities from 30 to 500 km $s^{-1}$. They also
illustrated that the times when the downflow motions started to be
seen corresponded to the times when bursts of non-thermal emissions
in hard X-rays and microwaves were emitted. \citet{lin05} pointed
out an average outflow velocity ranged from 640 to 1075 km $s^{-1}$.
The velocities of the upflows and downflows in our paper are
45.4-556.7 km $s^{-1}$, consistent with the results mentioned above.
In section 3.3, we described an outflow of a bright cloud from the
reconnection site. The bright cloud is plasmoids which may be
accelerated by the tension force of newly reconnected magnetic field
lines.

All the observations of the magnetic reconnection signature of the
flare can be explained by schematic diagrams shown in Fig. 6. In
order to better compare these diagrams with the observations, we use
the same expressions of physical parameters mentioned in section 3.
The dashed lines L2 and L3 are deduced from the information of
magnetic field structures, flare ribbon evolutions and post-flare
loop dynamics. Before the flare, there are three main loops in this
AR marked as L1, L2 and L3 in Fig. 6a. Because of some disturbance,
two anti-parallel field lines L1 and L2 meet resulting in the
formation of a current sheet. Then part of L1 and L2 are broken and
reconnect, the flare begins. As a result of this reconnection, the
inflow F1 (shown by the thick solid arrow in Fig. 6a), the newly
formed post-flare loops PFLs1 (arrowed in Fig. 6b), and the upward
and downward propagating plasmoids DF and UF (displayed by the
hollow thick arrows in Fig. 6b) appear. Several minutes later,
another reconnection occurs between L1 and L3. As a result of this
reconnection, the inflow F2 (displayed by the thick solid arrows in
Fig. 6b), the newly formed post-flare loops PFLs2 (arrowed in Fig.
6c), and downflows and upflows of accelerated plasmoids DF and UF
(shown by the hollow thick arrows in Fig. 6c) take place. The outer
edges of L1 and L3 form a cusp-shaped structure CUSP (arrowed in
Fig. 6c) and an X-shaped structure. The tension force of the
reconnected magnetic field lines accelerates the plasmoids out of
the current sheet, then the outflow OF (marked by the thick solid in
Fig. 6c) appears.

\acknowledgements

The authors are indebted to the \emph{TRACE}, \emph{YOHKOH} and
\emph{SOHO}/\emph{MDI} teams for providing the data. The work is
supported by the National Natural Science Foundations of China
(G40890161, 10703007, 10603008, 40674081, and 10733020), the CAS
Project KJCX2-YW-T04, the National Basic Research Program of China
under grant G2006CB806303, and Young Researcher Grant of National
Astronomical Observatory, Chinese Academy of Sciences.

\clearpage

\begin{figure}
\epsscale{1.} \plotone{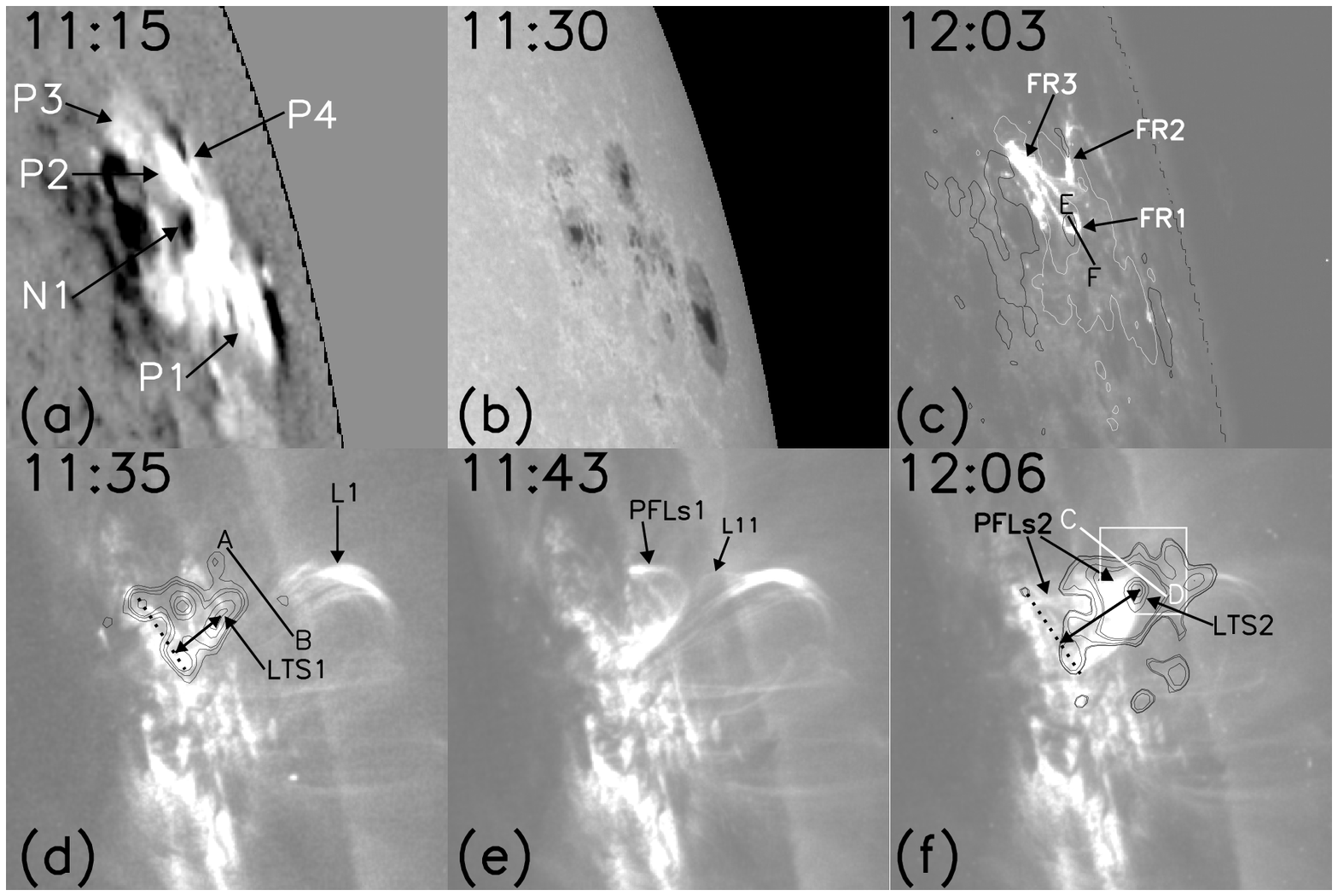} \caption{(a): A longitudinal
magnetogram observed by \emph{SOHO/MDI}. P1-P4 represent the
positive magnetic fields; and N1, the negative magnetic field.
(b-c): \emph{TRACE} WL and 1600 \AA~ images. FR1-FR3 are the flare
ribbons. (d-f): TRACE 195 \AA~images showing the evolution of EUV
loops. PFLs1 and PFLs2 denote the post-flare loops, L1 and L11, the
EUV loops, LTS1 and LTS2, the loop-top hard X-ray sources. The white
window in (f) represents the field of view (FOV) in Fig. 2. The
solid lines AB, CD, EF in (c), (d) and (f) show the position for
time slice evolution shown in Fig. 3. The contours in (c) show the
magnetic fields of the active region, while in (d) and (f), the hard
X-ray emission observed by \emph{YOHKOH/HXT}. The dotted lines in
(d) and (f) connect two footpoint hard X-ray sources, and the
two-head arrows represent the distance from the dotted lines to the
loop-top sources. The FOV is 200\arcsec $\times$ 200\arcsec.
\label{f1}}
\end{figure}

\begin{figure}
\epsscale{1.} \plotone{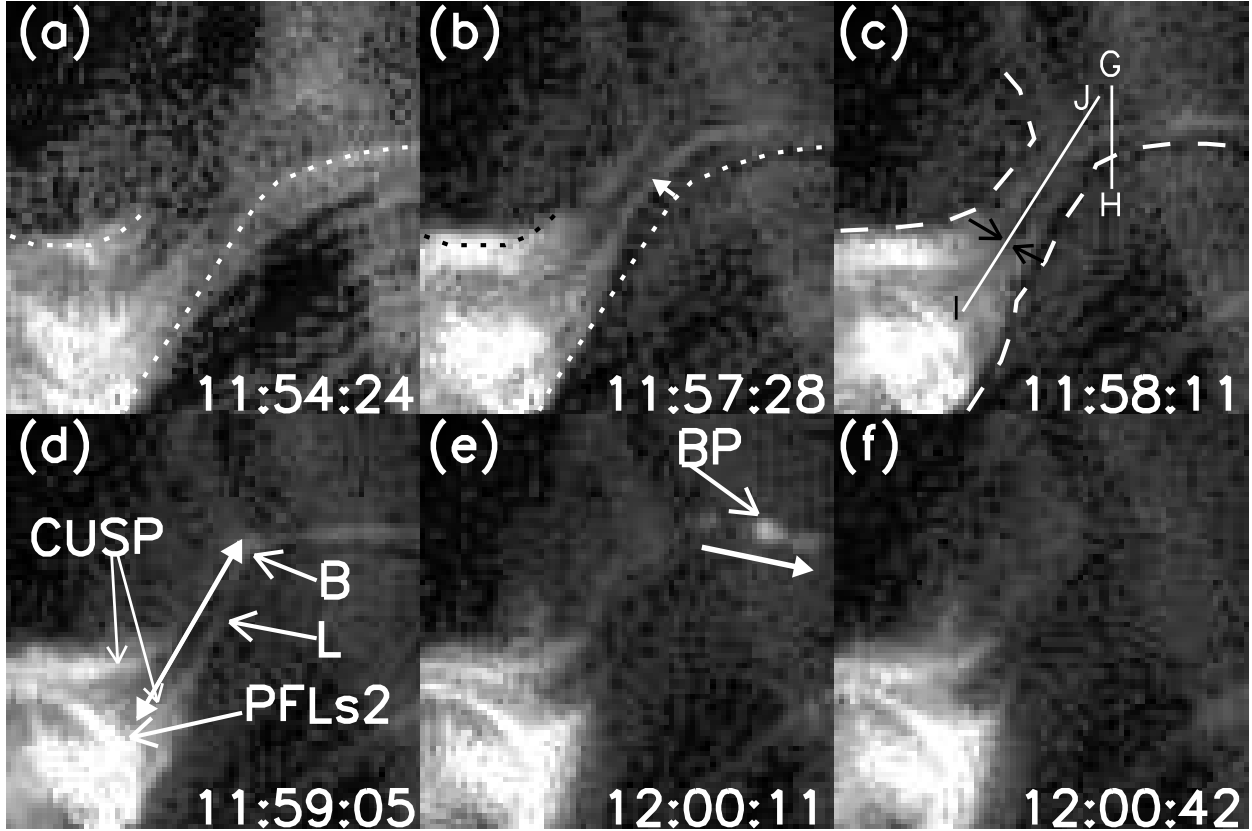} \caption{Time sequence of
\emph{TRACE} 195 \AA~images showing the evolution of the loops. The
dotted lines in (a) and (b) represent the loops at 11:54:24 UT, as
well as the dashed lines in (c), the out edges of loops at 11:58:11
UT. The solid arrow in (b) represents the moving direction of the
loops, as well as the solid arrow in (e), the moving direction of
the bright point. The two black arrows in (c) display the distances
between two EUV loops, and the solid two-head arrows in (d), the
length of EUV loops. CUSP means the cusp-shaped structure, PFLs2,
the post-flare loops, L, EUV loop, B, break point, BP, bright point.
The white solid lines GH and IJ in (c) show the position for the
distributions of several physical quantities displayed in Fig. 5.
The FOV is 40\arcsec $\times$ 40\arcsec. \label{f2}}
\end{figure}

\begin{figure}
\epsscale{0.5} \plotone{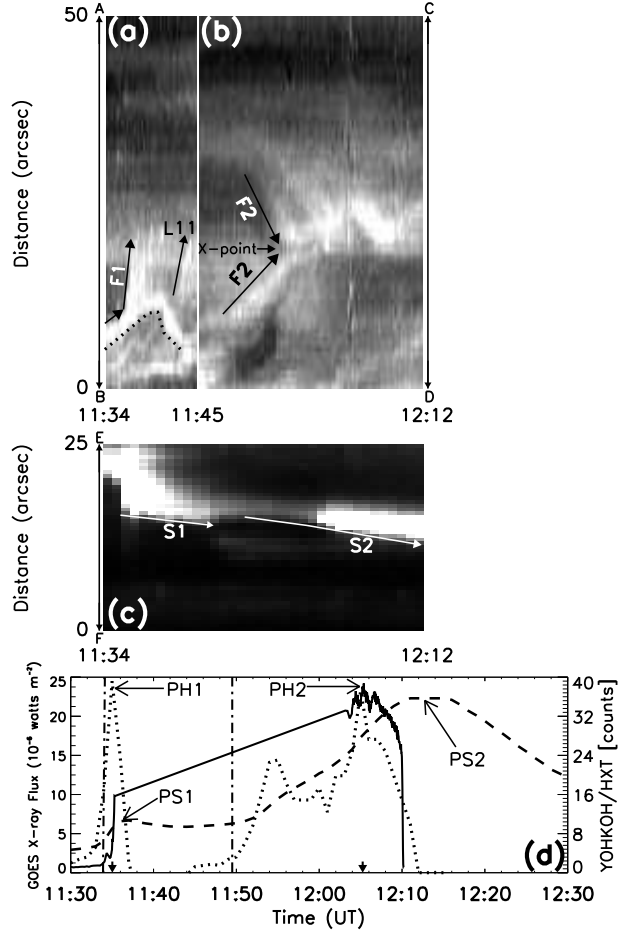} \caption{(a)-(c): Time evolution of
the one-dimension distribution of \emph{TRACE} 195 \AA~and 1600
\AA~intensity along the solid lines AB, CD, EF specified in Fig. 1,
respectively. F1 and F2 represent the sideward motions of the EUV
loops, as well as S1 and S2, the separations of FR1. The dotted line
in (a) outlines the out edges of loops. (d): The time integrated
hard X-ray flux (solid curves) from YOHKOH/HXT, the \emph{GOES}-10
1-8 \AA~soft X-ray flux (dashed curves) and their time derivative
(dotted curves). PS1-PS2 and PH1-PH2 represent the peaks of these
flux. The vertical dash-dotted lines represent the beginning times
of the sideward motions, and the two vertical solid arrows, the
times when we obtained hard X-ray images (see the contours in Figs.
1d and 1f). \label{f3}}
\end{figure}

\begin{figure}
\epsscale{1.} \plotone{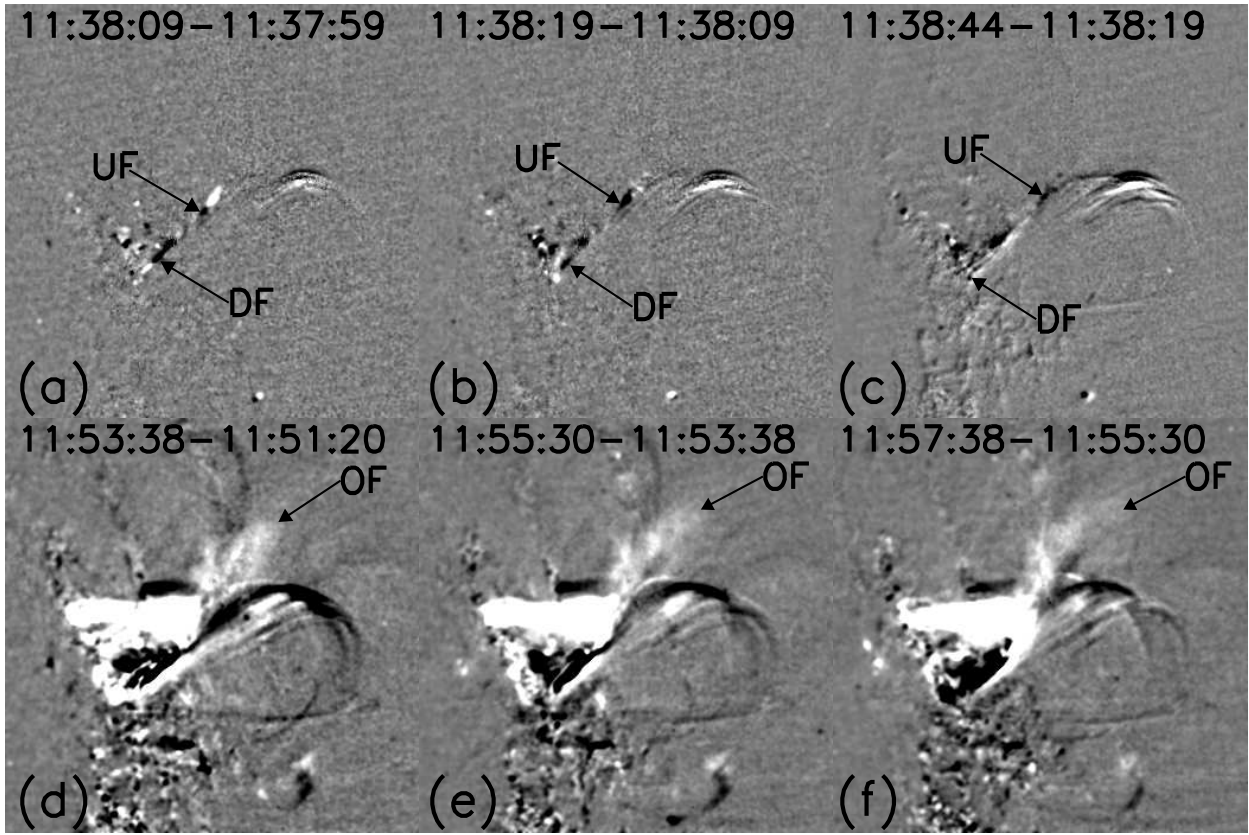} \caption{Time sequence of running
difference images observed by \emph{TRACE} 195 \AA~showing the
motions of plasmoids upward flow (UF) and downward flow (DF) along
loops (a-c), and outflow (OF) of plasmoids (d-f). The FOV is
175\arcsec $\times$ 175\arcsec. \label{f4}}
\end{figure}

\begin{figure}
\epsscale{0.8} \plotone{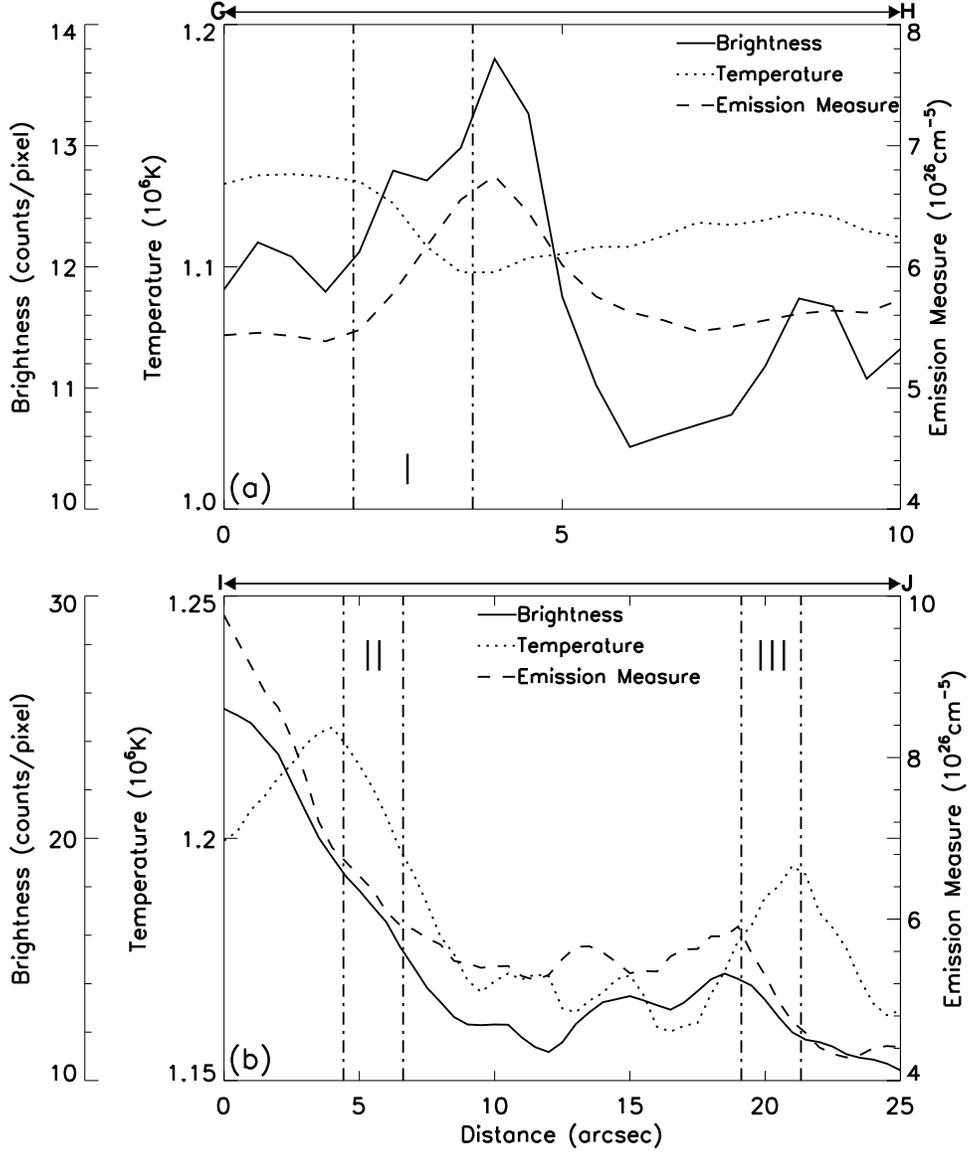} \caption{The distributions of
brightness (solid curves), temperature (dotted curves) and emission
measure (dashed curves) along the white solid lines GH (a) and IJ
(b) shown in Fig. 2c. The dash-dotted lines in (a) and (b) outline
the discontinuous regions ``I", ``II" and ``III". \label{f5}}
\end{figure}

\begin{figure}
\epsscale{0.6} \plotone{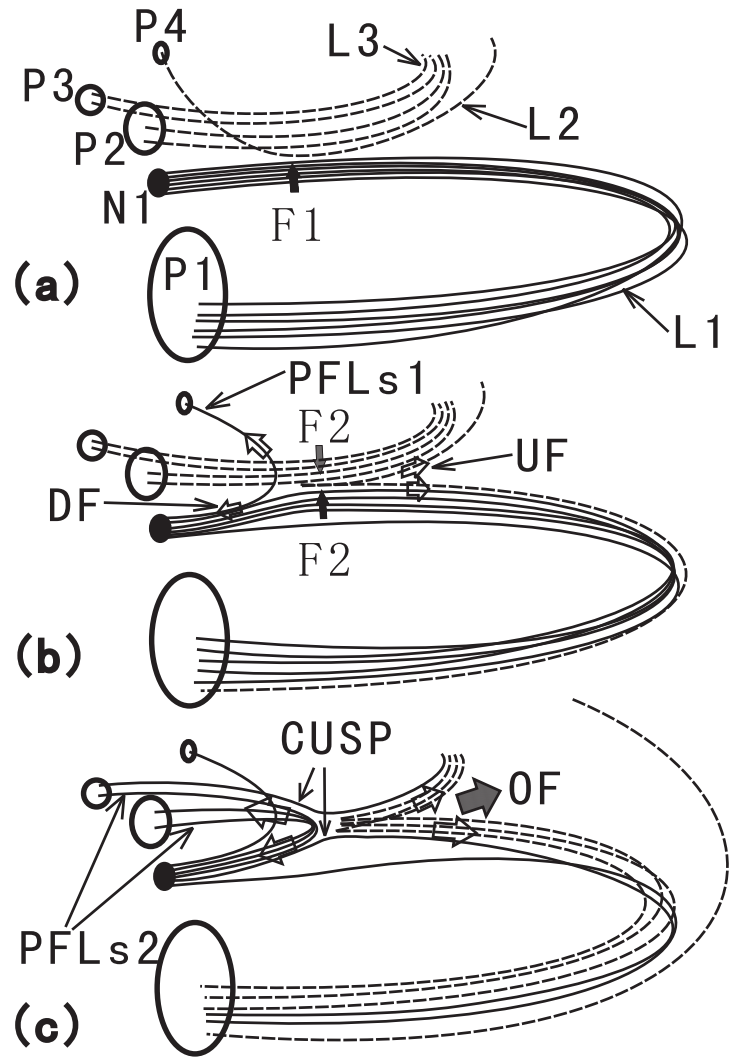} \caption{The schematic diagrams
illustrating the evolution of the flare event on 2000, March 23.
The ellipses represent the sunspots in the active region. The
lines (L1-L3) show the loops, and PFLs1 and PFLs2, the post-flare
loops. CUSP shows the cusp-shaped structure. The solid thick
arrows in (a) and (b) display the directions of the sideward
motions, as well as the hollow thick arrows in (b) and (c), upflow
(UF) and downflow (DF) of plasmoids. The solid arrow in (c) shows
the outflow (OF) of plasmoids. \label{f6}}
\end{figure}

\end{document}